\documentclass[a4paper]{llncs}
\usepackage[utf8]{inputenc}
\usepackage{amsmath} 
\usepackage[pdftex]{graphicx}
\usepackage{tabularx}
\usepackage{multirow}
\usepackage{comment}
\usepackage{calc}
\usepackage{pslatex}
\usepackage{listings}
\usepackage{url}
\usepackage{caption}
\usepackage{subcaption}
\newcommand{\alternativa}[1]{}
\newcommand{\ie}{\textit{i.e.},\ }
\newcommand{\eg}{\textit{e.g.},\ }
\newcommand{\etal}{\textit{et al.}\ }

\newcommand{\cost}{\mbox{cost}}
\newcommand{\length}{\mbox{length}}
\newcommand{\disj}{\mbox{disj}}
\newcommand{\ocorr}{\mbox{occurrences}}
\newcommand{\penal}{\mbox{penalty}}
\newcommand{\sharing}{\mbox{sharing}}
\newcommand{\selec}{\mbox{selecPaths}}
\newcommand{\best}{\mbox{bestSubset}}
\newcommand{\compat}{\mbox{compat}}
\newcommand{\potencial}{\mbox{potAgreg}}

\begin{document}

\title{Relieving Core Routers from Dynamic Routing with
off-the-shelf Equipment and Protocols}
\author{Margarida Mamede and Jos\'{e} Legatheaux Martins and Jo\~{a}o Horta}

\tocauthor{}

\institute{Technical Report \--- February 2016 \\
	   NOVA-LINCS, Departamento de Informática\\
           Faculdade de Ciências e Tecnologia, Universidade Nova de Lisboa\\ 
           2829--516 Caparica, Portugal\\
           \email{mm@fct.unl.pt, jose.legatheaux@fct.unl.pt, jp.horta@campus.fct.unl.pt} \\
           \url{http://nova-lincs.di.fct.unl.pt}, \url{http://di.fct.unl.pt} 
           }


\maketitle              

 
\begin{abstract}
To answer traffic engineering goals,
current backbone networks use expensive and sophisticated
equipments, that run distributed algorithms to implement dynamic 
multi-path routing (\eg MPLS tunnels and
dynamic trunk rerouting).
We think that the same goals can be fulfilled using a simpler approach,
where the core of the backbone only implements many a priori
computed paths, and most adaptation to traffic engineering goals only takes
place at the edge of the network. 
In the vein of Software Defined Networking, edge adaptation should be driven
by a logically centralized controller that leverages the available
paths to adapt traffic load balancing to the current demands and
network status.

In this article we present two algorithms to help building this vision. The first one selects
sets of paths able to support future load balancing needs and
adaptation to network faults. As the total number of required paths 
is very important, and their continuous availability  requires many FIB entries in
core routers, we also present a second algorithm
that aggregates these paths in a reduced number of trees.
This second algorithm achieves better results than
previously proposed algorithms for path aggregation.
To conclude, we show that  off-the-shelf equipment supporting
simple protocols may be used to implement routing with 
these trees, what shows that simplicity in the core can be
achieved by using only trivially available protocols and their most
common and unsophisticated implementations.

\end{abstract}

\section{Introduction}

Due to the endless explosion of video traffic distribution, social networks and
data center based content distribution, provider networks are subject to continuous
capacity and cost increases. In fact, the Internet and
its constituent networks have increased in scope and capacity 
many orders of magnitude in recent years,
and there are no signs of mitigation of this tendency in the near future \cite{cisco-forecast}.
This trend, as well as the competition among providers, including the competition among content
and network providers, drives a race to network optimization 
and to the simplification of its operations. Simple and sound engineering
principles have always been at the heart of the Internet success, scale and flexibility.
The motivation of this article is the need for simplicity, flexibility 
as well as optimization of routing and routers operations in a backbone network.

Routing in a network can be based on shortest path routing, 
a simple and greedy strategy that falls short when the goal is optimization.
When the goal is capacity usage optimization
while simultaneously providing good quality of service, one has to resort
to  traffic engineering  \cite{te-survey}.
Due to the popularity of  MPLS to support customer VPNs (Virtual Private Networks), 
and its ability to support  from fine-grained traffic controls to carrier grade requirements, 
the most common
traffic engineering solutions use multi-path 
routing mechanisms implemented
with LSPs (Label Switching Paths) \cite{Rosen2001}
or other types of tunnels.

The traffic optimization problem has been for long extensively studied,
specially in the offline case \cite{te-survey,heckmann2006}.
However, offline methods, in general, do not deal with network faults,
which are more common 
than desirable \cite{failures2008}, and traffic demands can also
abruptly change due to traffic bursts, which may happen in shorter periods than 
the average traffic engineering cycles (\ie offline load distributions are periodically adjusted
in face of the availability of new traffic demand estimates).

Dynamic online traffic engineering solutions are more complex  and
sometimes suboptimal, since truly optimal solutions require a global and consistent view
of the tunnels load.
Often, these solutions  rely on distributed flooding algorithms
allowing each node to globally estimate tunnels load and network status. 
When traffic demands change, or faults occur, 
routers asynchronously compute new paths and load distributions,
sometimes conflicting with the optimal goals \cite{mpls-te-latency}.
Due to the complexity  and sophistication of their software and algorithms,
core routers supporting all these features are much more expensive
than simple switches of similar capacity.

The limitations of the offline traffic engineering solutions,
 the complexity, limitations and cost of the dynamic ones, 
 the endless growth of the forecasted traffic demands, and
the popularity of data centres with many thousands machines
are significant motivations to search for simpler, more effective and less 
expensive ways of network configuration, traffic optimization and management, 
as well as less expensive network gear.
 Moreover, the need for traffic optimization is now common to several other
 types of networks besides traditional ISPs networks. This requirement is now
common to intra data center \cite{vl2,Mudigonda2010,past} and inter data center 
networks \cite{b4-sdn-wan,ms-sdn-wan}.

This state of affairs needs simpler and more effective 
engineering solutions and core routers
shielded from the complexity of dynamic route (re)computation as proposed in \cite{drrch}. 
This is in part the same direction
taken by the so called  Software Defined Networking approach (SDN)
that aims at \emph{``separating routing from routers"} \cite{OpenFlow,road-to-sdn,sdn-internet}.
Some recent publications report traffic
engineering experiences, \eg \cite{Suchara2011,b4-sdn-wan,ms-sdn-wan}, that go 
in that direction. All share part or most of the following tenets:

\begin{itemize}
\item keep core routers as simple as possible and concentrated on forwarding; their main purpose is to 
make available the different needed paths; when faults occur, they only
report their occurrence and let 
the edge (\eg edge routers, controllers
 or servers in data centres) adapt to the new situation \cite{Suchara2011};
\item concentrate any required in-network complexity in edge devices; 
they are in charge of load balancing incoming flows
across the available paths; their flow distribution policies are pre-computed \cite{Suchara2011}
or are controlled by logically centralized controllers \cite{b4-sdn-wan,ms-sdn-wan};
\item execute traffic optimization algorithms offline \cite{Suchara2011} or by some logically centralized network controller \cite{b4-sdn-wan,ms-sdn-wan}
that is aware of the global network status and  traffic demands.
\end{itemize}

In this paper we adhere to this vision and contribute with two algorithms to help the
first requirement: setting up a generic off the shelf core, shielded from edge 
or controller dynamic decisions.

We consider that, due to scalability issues, edge flows are not
individually routed \cite{sdn-internet}. Instead, these flows are partitioned into sets, sometimes
called trunks, and routed  through tunnels. This approach is frequent in large backbones and 
interconnection networks. In the rest of the paper the terms path and tunnel will
be used interchangeably.

In the above vision, core routers must provide a set of edge-to-edge paths
suitable for supporting the needed routing strategies and load distributions, as
implied by optimization computations. The first algorithm
presented in this paper deals with the a priori computation of theses paths.

%

The required number of paths is very important in large networks, since in a
$n$ ingress/egress node  network, using $k$ different paths for
each edge node pair, $O(k n^2)$ paths are needed. Additionally, as their setup should be
possible with off the shelf equipment and known protocols, we also propose an algorithm
for aggregating these paths into a reduced number of trees, and show that these routing trees
are easily supported with affordable equipment and by known protocols. 
This second algorithm is a path aggregation one,
which is also useful to tackle the general problem of path aggregation
into trees. Trees are a simple way to reduce the routing
state needed in routers and are well supported by many off the shelf 
switching equipment.

As discussed in the corresponding sections,
both algorithms achieve their goals in better ways than previous similar algorithms presented in the 
literature, and were extensively tested and compared using several types of networks.

We begin by presenting, in section \ref{sec:networks}, the notation used
and the networks selected for
the algorithms evaluation. These encompass a set of synthetic networks as well as
several real world backbone models. 
The next two sections are devoted to the
algorithms. Both sections have the same structure: algorithm goals, related work, the proposed algorithm and
the evaluation. 
We proceed with a discussion on the usage of the algorithms and 
include a survey on how multi-path routing based on a set of trees
can be implemented with off the shelf equipment and different types
of technologies: MPLS, learning switches supporting flooding
and static VLANs, routers supporting longest prefix matching and static routes, 
and OpenFlow \cite{OpenFlow}. 
Finally, we conclude the article in the discussion and conclusions section.

\section{Notation and networks used for evaluation purposes}
\label{sec:networks}

\subsection{Graphs characterization and notation used in the algorithms}
\label{sec:notations}

During the presentation and evaluation of the algorithms we use 
networks modelled as graphs. We start this section by briefly and 
informally reviewing  the class of graphs we deal with.

A network is modelled by a graph $G=(V,E)$ that is
\textit{undirected} (so $(v_1,v_2)$ and $(v_2,v_1)$ denote the same edge),
\textit{simple} (there are no loops nor parallel edges),
\textit{connected} (there is a path between any two nodes),
and 
\textit{weighted}, being the weight of each edge strictly positive.
Additionally, 
let $N \subseteq V$ be the set of 
edge nodes originating and terminating traffic,
and $n=|N|$. 
For all pairs $(x,y)\in N^2$, with $x < y$ (for any total order on $N$), 
we want to compute $k>0$ distinct paths from $x$ to $y$ 
in order to support multi-path routing 
(see section~\ref{sec:selection}).
The set $S$ of the computed paths has cardinality 
$|S| \approx k\frac{n(n-1)}{2}$ 
because,
for some pairs of nodes, it is impossible to find $k$ different paths and,
in some other cases, more than $k$ paths will be selected 
(see section~\ref{sec:selection-evaluation}).  

The computed paths will be aggregated into a reduced set $T$ of trees 
covering them (see section~\ref{sec:trees}), such that $|T| \ll |S|$. 
A \textit{tree} $t$ is an undirected simple graph that is
connected and acyclic, and
it \textit{covers} a path $p$ if $p$ is a path in $t$.

To test the algorithms, and for comparison purposes, 13 networks were selected.
These networks are of two types: 
\textit{synthetic regular networks} 
with a priori known best paths and 
for which it is possible to deduce the minimum number of covering trees, 
which are used for algorithm evaluation purposes,
and a set of representative \textit{backbone networks} 
whose best paths and corresponding optimal covering tree sets are unknown.

\subsection{Synthetic regular networks}
\label{sec:regular-networks}

Four synthetic regular networks configurations are used:
full mesh, ring, hierarchical and folded clos. 
In all of them, the weight of any edge is $1$,
that is, all links have cost $1$.
We will characterize the \textit{best paths} in these networks,
which are the paths that should be computed by a path selection algorithm.

\begin{figure}
\centering
\begin{subfigure}{.45\textwidth}
  \centering
  \includegraphics[width=3cm]{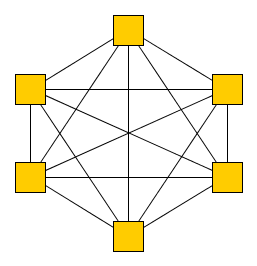}
  \caption{full mesh}
  \label{fig:fullmesh}
\end{subfigure}%
\begin{subfigure}{.45\textwidth}
  \centering
  \includegraphics[width=3cm]{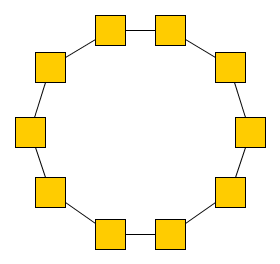}
  \caption{ring}
  \label{fig:ring}
\end{subfigure}
\caption{Examples of full mesh and ring networks}
\end{figure}

In full mesh and ring networks all nodes are origin and destination of
traffic (therefore $N=V$), see section \ref{sec:tree-routing}. 
A $n$-node full mesh is a clique 
(as illustrated in Figure~\ref{fig:fullmesh}).
Between any two distinct nodes,
there is a shortest path (with cost 1) and $n-2$ paths of cost 2.
These are the best paths,
leading to a set with $(n-1) \frac{n(n-1)}{2}$ best paths. 
In a ring network each node has degree 2 and there are
two (disjoint) simple paths to reach any other node 
(see Figure~\ref{fig:ring}).
The set of best paths is the set of all simple paths and
has $2 \frac{n(n-1)}{2}$ elements.
In both configurations,
the best paths can be aggregated into $n$ trees, 
each one rooted at the origin node.

\begin{figure}
\centering\begin{subfigure}{.45\textwidth}
  \centering
  \includegraphics[width=4cm]{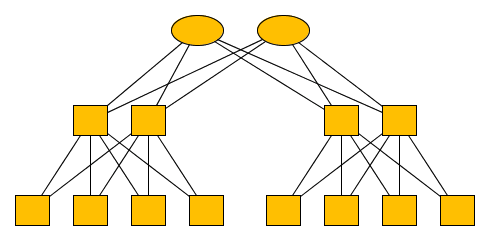}
  \caption{A two level hierarchical}
  \label{fig:fattree}
\end{subfigure}%
\begin{subfigure}{.45\textwidth}
  \centering
  \includegraphics[width=4cm]{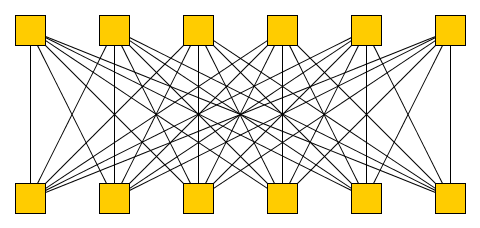}
  \caption{folded clos}
  \label{fig:foldedclos}
\end{subfigure}
\caption{Examples of hierarchical (tree-like) and folded clos networks}
\end{figure}

In traditional small to medium data center networks,
hierarchical tree-like networks are commonly used,
see Figure~\ref{fig:fattree}.
In these hierarchical networks, the best
paths between any two leaf
nodes (the only ones that belong to $N$)
are the shortest ones and there are $2^{2m-1}$ such paths, 
where $m$ is the minimum number of levels that must be climbed to reach the
destination node (all these paths are \emph{valley-free})
\footnote{A valley-free path in this context can be seen as
a path that ``climbs and descends only once".}.

A folded clos network is defined by two distinct layers, 
forming a bipartite graph, 
where each node of the lower layer (which corresponds to the set $N$) is
directly connected to every node of the upper layer.
Figure~\ref{fig:foldedclos} presents a
folded clos network with the same number of nodes in each layer. 
For each pair of nodes of the lower layer,
the best paths are the shortest paths, 
which are valley-free and pairwise disjoint.
Their number is equal to the size of the upper layer. 
Therefore, in a folded clos network with $n$ nodes in each layer, 
there are $n\frac{n(n-1)}{2}$ best paths, 
which can be aggregated into $n$ trees. 
Those networks are also traditionally used in data centres 
and have the interesting property
of maximizing path disjointness among nodes. 

In the first 6 rows of Table~\ref{tab:networks} 
the used regular networks are characterized in terms of 
number of nodes, 
number of edges, 
number of best paths, and 
minimum number of trees to cover the best paths.

\begin{table}
\caption{Characterization of the networks used in the evaluation 
(``\---" stands for unknown)}
\label{tab:networks}
\begin{center}
\begin{tabular}{ |c|c|c|c|c|c|c| }
		\hline
		\multirow{2}{8ex}{Network} 
		& \ \# nodes \ 
		& \ \# edges \ 
		& \ n  = \ 
		& \ \# best \ 
		& min
		& Characterization \\
		& $| V |$ 
		& $| E |$ 
		& $| N |$ 
		& paths 
		& \ \# trees \
		& (as used in this paper) \\ 
		\hline
		full mesh & 12 & 66 & 12 & 726 & 12  &  A clique graph \\
		ring & 12 & 12 & 12 &  132 &  12 & A circular graph  \\
		hierarchical 2 & 14 & 24 & 8  & 152 &  8 & A two level tree-like graph \\
		hierarchical   3 & 30 & 56 & 16 & 2352 &  32 &  A three level tree-like graph \\
		folded clos 6    & 12 & 36 & 6 & 90 & 6  & A bipartite graph \\
		\ folded clos 12 \ & 24 & 144  & 12 & 792  & 12  & A bipartite graph \\
		
		Abovenet & 15 & 44 & 15 & \---  & \--- & USA backbone  \\
		ATT & 35 & 68 & 35 & \---  & \--- & USA backbone  \\
		B4 & 12 & 19 & 12 & \---     & \--- & World wide DC backbone  \\
		Géant & 32 & 49 & 32 & \---  & \--- & \ European research backbone \ \\
		NTT & 27 & 63 & 27 & \---  & \--- & World wide backbone  \\
		Sprint & 32 & 64 & 32 & \---  & \--- & World wide backbone  \\
		Tiscali & 30 & 76 & 30 & \---  & \--- & European backbone  \\
		\hline
\end{tabular}
\end{center}
\end{table}

\subsection{Backbone networks}
\label{sec:backbone-networks}

Backbone networks used for testing purposes were selected 
using several criteria. 
First, only POP level network models were used
since routing policies are more relevant at this level than at routers level. 
Besides, 
using the routing concretizations presented in section \ref{sec:lpm}, 
several routers of the same POP can be collapsed in one logical one and ECMP 
(Equal Cost Multi-Path Routing) can be used to load balance 
traffic across several \emph{parallel links} directly connecting two POPs. 
Second,  public depicted backbones were preferred 
to avoid having to resort to synthesized  randomized
backbone  topologies. 

One education and research backbone (Géant), 
a world wide backbone (NTT Communications) and 
the network used in \cite{b4-sdn-wan} 
were chosen since their
configurations are publicly sketched.
These three backbones were approximately 
mapped from their publicly presented diagrams. 
Additionally, 
we used four commercial ISP backbones mapped by the Rocketfuel project 
\cite{rocketfuel}:
Abovenet, ATT, Sprint and Tiscali. 
Their mappings are more than 10 years old, 
but we still deem them representative of the most important 
topological characteristics of real wide area backbones.

In all those backbone networks, 
as the link capacities were, in general, not publicized,  
latency 
(approximately inferred from the geographic distance 
between the cities where the POPs are located) 
is used as the cost metric. 
This metric is closely related to latency inflation 
which is particularly relevant for the tested algorithms 
when applied to wide area backbones. 
To speedup computations, 
backbone graphs were (iteratively) shrunk  
by pruning them from nodes of degree 1, 
since these nodes do not introduce any extra diversity or path alternatives. 
The last 7 rows of Table \ref{tab:networks} present the characteristics
of the retained backbone networks.

The number of required paths and trees will increase 
when used to drive a real network routing configuration. 
Degree 1 nodes will be reinserted, 
what will increase the number of paths, but not the number of trees.
Besides, provisioning of different 
classes of service to customers will also increase the number of paths
and trees by a factor proportional 
to the number of different traffic classes used. 
However, all multi-path routing methods discussed in section \ref{sec:tree-routing} 
require the same increase in backbone state and control complexity if several service classes are used. 
Resource reservation for traffic classes 
(and therefore for different trees) is outside the scope of this paper.
Anyway, in large backbones, 
shaping and admission control, when applied, are performed at the 
edge to relieve core routers of these concerns.


\section{Path selection}
\label{sec:selection}
\subsection{The problem and known solutions}

Given a network and a traffic matrix, to find a
multi-path continuous traffic distribution that minimizes the load in all the network links
is a well known optimization problem 
\cite{te-survey,heckmann2006}.
This traffic distribution implicitly elects a set of paths used for routing. With
several matrices, the union of the sets of paths will eventually stabilize.

Practice has shown that it is useless to select more than a small
number of paths between each pair of edge nodes. In fact, the problem 
may be approached in the reverse direction: 
to speed up the optimization computation, one can a priori restrict the traffic
distribution to a subset of the available paths, computed in a traffic
matrix independent way, see chapter 12 of \cite{heckmann2006} for example. 
The common reported experience
(\eg \cite{b4-sdn-wan}) shows that restricting the traffic distribution to
 a subset of paths can speedup significantly
the optimization process, without any practical relevant impact in the
 final result optimality. In general, that subset is composed of 
 $k$ (almost) shortest paths.


The most common approach to tackle network faults, in the context of traffic
engineering, is to complete the set of paths used to optimally route traffic 
with backup paths. 
Some of the network faults may lead to non optimal traffic routing. 
In \cite{Suchara2011} a method is proposed to deal with this degradation. 
Given a set of the most common faulty scenarios, 
the optimal traffic distribution of the considered traffic matrix is computed for each one. 
Using the paths computed for each scenario,
it is possible to compute a global subset of the most interesting ones. Finally, for each
faulty scenario, an optimal traffic distribution is computed over the selected paths.
The method requires several a priori collected pieces of information: 
traffic matrices and the common faulty scenarios. 
Both are hard to find but when
the network managers have years of accumulated traffic and fault statistics
with their network.
Moreover, in what concerns network faults, most of them stem from old
components with intermittent behaviour, software bugs and planned reconfigurations
\cite{failures2008}. Additionally, estimating a meaningful traffic matrix is also
a non trivial problem \cite{te-survey}.

Due to these difficulties, 
it is common to resort to heuristic methods to a priori 
compute a suboptimal set of paths. 
A simple example, described in \cite{lsps}, 
consists in selecting, for each pair of edge nodes, a set of paths, in general just two, 
that includes at least two node disjoint paths 
whose cost stretch over the shortest ones is constrained. 
When it is not possible to find such a pair of paths, 
a shortest path and a disjoint path in terms of nodes with it are chosen. 
The authors use a heuristic that privileges fault-tolerance over/above traffic distribution.

%

SPAIN is a multi-path data center routing proposal \cite{Mudigonda2010} 
which also uses,
for each pair of edge nodes, 
a set of paths computed independently of a traffic matrix. 
The algorithm has as input a network graph,
a pair of edge nodes (origin and destination) and 
the number $k$ of paths from the origin to the destination to be computed.
In each step, 
the algorithm starts by computing a shortest path $p$.
Then, the cost of every edge of $p$ is increased
by a large number (such as the sum of the initial costs of all graph edges),
to prevent its use in the next iteration if cheaper alternatives are available.
The algorithm terminates
when $k$ different paths have been found
or the path computed in some iteration has already been selected in a previous step.
This algorithm also privileges edge disjointness over/above traffic distribution
but, as we will see later, it is often unable to compute $k$ paths
even when they exist.

We also use a heuristic to select the best paths to support multi-path
routing in a traffic matrix and fault agnostic way. 
However, we try to reconcile both goals: 
to select a set of paths deemed good for traffic distribution, 
as well as being able to cope with network faults. 
Moreover, since we also target backbone networks, 
we should limit latency inflation of the selected paths.

Another common traffic matrix agnostic method 
of computing paths for traffic load balancing 
is used by the two phase multi-path routing scheme, 
inspired by the well-known
Valiant Load Balancing (VLB)  \cite{valiant1982scheme,hose-model} method.
To route from node $x$ to node $y$ with two phase routing, 
all paths formed using a shortest path from $x$ to an intermediate node $i$,  
followed by a shortest path from $i$ to $y$ 
are used to send traffic from $x$ to $y$.\footnote{The shortest paths 
from $x$ to $y$ are included because $x$ (and $y$) can be the intermediate node.}
The resulting paths provide a good solution to traffic load balance 
in data center networks where path latency differences are not very relevant, 
but are not so suited to WAN backbones spanning one or more continents. 
Besides, the method does not guarantee any disjointness degree 
among the selected paths.

%

\subsection{Algorithm 1 -- the path selection algorithm}
\label{sec:algoritmo1}

Algorithm~1 selects a set of paths in a network $G=(V,E)$
(which has been characterized in section~\ref{sec:notations})
by computing,
for every pair of distinct edge nodes, 
$(x,y) \in N^2$, with $x<y$,
a set of paths from $x$ to $y$.
So, let $(x,y)$ be the pair of nodes for which we intend
to compute $k>0$ distinct paths.
Basically, 
those paths should be ``short'' and ``share'' few edges.
The next definitions are needed to formalize these notions.

For every path $p$ in $G$,
the \textit{length} of $p$ (denoted by $\length(p)$)
is the number of its edges and
the \textit{cost} of $p$ ($\cost(p)$)
is the sum of the weights of its edges.
Let $\mathcal{P}_{xy}$ be the set of simple paths from $x$ to $y$ in $G$
(which is not empty because $G$ is connected) and 
$\mathcal{C}_{xy} \subseteq \mathcal{P}_{xy}$ be the set of 
\textit{min-cost} paths:
$$\mathcal{C}_{xy} =
  \{p \in \mathcal{P}_{xy} \mid 
  (\forall p'\in \mathcal{P}_{xy})\; \cost(p) \leq \cost(p') \}.
$$
An \textit{optimal} path $o$ is a min-cost path that has the smallest length,
that is, 
$$o \in \mathcal{C}_{xy} \;\; \mbox{ and } \;\;
  (\forall p\in \mathcal{C}_{xy})\; \length(o) \leq \length(p).
$$

A set of \textit{pairwise disjoint paths} (in terms of edges) is 
a set of paths 
where any edge of the graph occurs at most once.
Given a set $S$ of paths,
the \textit{disjointness degree} of $S$ ($\disj(S)$) is 
the maximum number of pairwise disjoint paths in $S$ 
or, more precisely,
the maximum cardinality of the subsets of $S$ 
whose paths are pairwise disjoint.
For example, the disjointness degree of 
$A=\{1\, 2\, 4,         \,
     1\, 2\, 3\, 5\, 4, \, 
     1\, 2\, 6\, 4,     \, 
     1\, 3\, 2\, 4,     \,
     1\, 5\, 4          \}$
is three because paths in
  $\{1\, 2\, 6\, 4, \, 
     1\, 3\, 2\, 4,     \,
     1\, 5\, 4          \}$
are pairwise disjoint and, 
since there are three paths in $A$ that contain edge $(1,2)$, 
only one of them can belong to a set of pairwise disjoint paths.

Sets with higher disjointness degree are usually more fault-tolerant.
For this reason, 
the algorithm selects a set that maximizes disjointness.
Among the sets with maximum disjointness degree, 
it computes one that minimizes ``edge sharing'',
``penalizing'' the repeated use of the same edges.
Function $\penal(S,e)$ assigns a penalization to 
the (possible) use of edge $e$ in the set of paths $S$, 
where  $\ocorr(S,e)$ denotes the number of paths in $S$ that contain $e$:
$$\penal(S,e) = 
\left \{ \begin{array}{l@{\;\;\;}l}
         0,                   & \mbox{if } \ocorr(S,e) \leq 1;        \\
	 (|S|+1)^{\ocorr(S,e)}, & \mbox{if } \ocorr(S,e) \geq 2.
	\end{array}
\right .
$$
The \textit{edge sharing} in $S$ is the sum of 
the penalizations of the use of the graph edges in $S$: 
$$\sharing(S) = \sum_{e \in E} \penal(S,e).$$
For the sake of illustration, 
let us compare the following two subsets of $A$ of cardinality 4 
and disjointness degree 3: 
$$B_1=\{1\, 2\, 4,         \,
        1\, 2\, 6\, 4,     \, 
        1\, 3\, 2\, 4,     \,
        1\, 5\, 4          \} \;\; \mbox{ and } \;\;
  B_2=\{1\, 2\, 3\, 5\, 4, \, 
       1\, 2\, 6\, 4,     \, 
       1\, 3\, 2\, 4,     \,
       1\, 5\, 4          \}.$$
As $\sharing(B_1)=2 \times 5^2$ 
(because the two edges of path $1\, 2\, 4$ occur twice in $B_1$)
and $\sharing(B_2)=3 \times 5^2$
(because $(1,2)$, $(2,3)$ and $(5,4)$ occur twice in $B_2$),
set $B_1$ is considered to be better than $B_2$.

\begin{figure}
\normalsize
\centering
\fbox{\rule[-65pt]{0pt}{135pt}
$\begin{array}{l@{\;\;\;\;\;\;}l}
\multicolumn{2}{l}{\selec(G,x,y,k,h,f) \; =}            \smallskip \\
& \left \{ \begin{array}{l@{\;\;\;}l}
  \best(\mathcal{C}_{xy}, k, \emptyset),               
  & \mbox{if } |\mathcal{C}_{xy}| \geq k;                \smallskip \\
  \mathcal{I}_{xy}^{hf},
  & \mbox{if } |\mathcal{C}_{xy}| < k \, \mbox{ and } \,
               |\mathcal{I}_{xy}^{hf}| \leq k;           \smallskip \\
  \mathcal{C}_{xy} \cup 
  \best(\mathcal{I}_{xy}^{hf} \setminus \mathcal{C}_{xy}, 
        k-|\mathcal{C}_{xy}|, \mathcal{C}_{xy}),             
  & \mbox{otherwise}.
\end{array}\right .

\bigskip\\
\multicolumn{2}{l}{\best(D,n,F) \mbox{ is a set } R 
  \mbox{\ that fulfils the following three properties:}} \smallskip \\
& \begin{array}{l@{\;\;\;}l}
  \bullet & R \in [D]_{n} \;\;\;\;\;\;\;
  (R \mbox{ is a subset of } D \mbox{ of cardinality } n);
                                                    \smallskip \\
  \bullet & (\forall X\in [D]_{n})\; \disj(R \cup F) \geq \disj(X \cup F);
                                                    \smallskip \\
  \bullet & (\forall X\in [D]_{n})\; 
       \disj(X \cup F) = \disj(R \cup F) \Rightarrow 
       \sharing(R \cup F) \leq \sharing(X \cup F).
\end{array}
\end{array}
$}
\caption{Set of selected paths from $x$ to $y$ in $G$,
for $k$ desired paths.
$\mathcal{C}_{xy}$ is the set of min-cost paths and 
$\mathcal{I}_{xy}^{hf}$ is the set of interesting paths
whose length and cost depend on $h$ and $f$, respectively.}
\label{fig:algSelec}
\end{figure}

Function $\selec$, presented in Figure~\ref{fig:algSelec},
characterizes the set of paths selected in $G$ from $x$ to $y$.
Apart from the intended number of paths ($k > 0$),
it has two more parameters, $h$ and $f$,
which allow us to restrict the length and the cost of the returned paths,
as it will be explained later.
If $U$ is a set and $n$ is an integer such that $0 \leq n \leq |U|$,
let $[U]_{n}$ denote the set of all subsets of $U$ with $n$ elements.

In the first case,
when there are at least $k$ min-cost paths,
a ``best'' subset of $\mathcal{C}_{xy}$ of cardinality $k$ is returned.
This means that the computed set $R$ satisfies the following properties:
\begin{itemize}
\item $R$ is a subset of $\mathcal{C}_{xy}$ of cardinality $k$:
      $R \in [\mathcal{C}_{xy}]_{k}$;
\item $R$ has the maximum disjointness degree:
      $(\forall X\in [\mathcal{C}_{xy}]_{k})\; \disj(R) \geq \disj(X)$;
\item Among the sets with maximum disjointness degree,
      $R$ has the minimum edge sharing:
      $(\forall X\in [\mathcal{C}_{xy}]_{k})\; 
       \disj(X) = \disj(R) \Rightarrow \sharing(R) \leq \sharing(X)$.
\end{itemize}

In the two other cases, 
it is necessary to consider paths with higher cost.
The selectable paths, named \textit{interesting} paths,
are min-cost or ``near-optimal'' paths.
This proximity is specified by two parameters, 
$h \geq 0$ e $f \geq 1$:
$h$ limits
the difference in length between a near-optimal and an optimal path,
whereas $f$ controls the cost of a near-optimal path,
which cannot exceed the minimum cost by a factor of $f$.
The set $\mathcal{I}_{xy}^{hf}$ of the \textit{interesting} paths 
from $x$ to $y$ is defined as follows,
where $o$ denotes an optimal path from $x$ to $y$:
$$\mathcal{I}_{xy}^{hf} = \mathcal{C}_{xy} \cup
\{p \in \mathcal{P}_{xy} \mid 
 \length(p) \leq \length(o) + h \; \wedge \;
 \cost(p) \leq f \times \cost(o)\}.
$$

If the number of interesting paths does not exceed $k$ 
(the second case of function $\selec$), 
the returned set is the set of interesting paths.
So it may have less paths than what has been asked for, 
in which situation 
the network manager should be warned to take corrective actions, 
if possible. 

When there are more than $k$ interesting paths
(the last case of $\selec$ and the most common one), 
the result set $S$ should include all min-cost paths 
as well as some other interesting ones, 
\ie it is a union of the form $S=\mathcal{C}_{xy} \cup R$,
where $R$ is a ``best'' subset of 
$\mathcal{I}_{xy}^{hf} \setminus \mathcal{C}_{xy}$ 
of cardinality $k-|\mathcal{C}_{xy}|$
taking into account the fixed part $\mathcal{C}_{xy}$.
As in the first case,
best means that the returned set $S$ has maximum disjointness degree and, 
among unions of that form of equal disjointness degree,
$S$ has minimum edge sharing.
It is easy to verify that both definitions of best subset can be merged,
as it is shown in Figure~\ref{fig:algSelec}.

In what concerns the implementation of the algorithm, 
the computation of the min-cost paths is performed efficiently 
by a variant of the Dijkstra algorithm \cite{dijkstra1959ntp}.
However,
near-optimal paths and best sets are computed by brute force methods,
which use data structures or orderings so as to reduce the search space.
Although the algorithm is not polynomial,
it computes the result in useful time in
the contexts for which it has been created 
(as we shall see in the next section). 
The number of paths between each pair of nodes ($k$) is as few as 3 or 5,
for example, and 
the number of interesting paths is controlled by parameters 
$h$ and $f$.

\subsection{Algorithm 1 evaluation}
\label{sec:selection-evaluation}

In all the evaluation results presented in this paper, 
the algorithms were implemented in Java, using Java VM version 8, 
in a sequential version, without
exploiting any parallelism of the supporting computer.
Programs were executed in a
2.4 GHz Intel i5 CPU with 8 GB of 1600 MHz DRAM.
Execution times will be referred as appropriate. 
Source code, input sets and results are all available
from the authors.

Given a concrete network graph $G$ and 
the values for parameters $k$, $h$ and $f$, 
Algorithm~1 was used to compute the paths between 
each pair of edge nodes $(x,y)$ such that $x < y$,
where the order relation on nodes is the order relation on their indexes.

\subsubsection*{Synthetic regular networks.}

The tests performed with these networks aimed at verifying 
if Algorithm~1 was able to select the best paths
(defined in Section~\ref{sec:regular-networks})
with an adequate choice of parameters.
Since all edges have cost 1, 
parameters $k$ and $h$ alone allow the selection of those paths, 
and parameter $f$ loses its significance
(any value $f \geq (h+1)$ neutralizes its effect).

The best paths in a $n$-node full mesh
are selected using $k = n-1$ and $h = 1$. 
In a ring network of $n$ nodes,
the best paths are selected using $k = 2$ and $h = n-2$.
For the hierarchical networks, 
we used $k = 2^{2m-1}$ and $h = 0$,
where $m$ denotes the level of the hierarchy.
Finally, 
the best paths in a folded clos network with $n$ nodes in each layer
are obtained with $k = n$ and $h = 0$.

Algorithm~1 computed the best paths for all regular networks. 
The SPAIN algorithm computed the same set of paths 
except with the hierarchical networks, 
when it is unable to select all the best paths 
between pairs of nodes that are connected by paths 
crossing the upper layer of the hierarchy. 
There are approximately 57\% and 80\% of such pairs 
in hierarchical networks with two and three levels, respectively. 

That behaviour is due to the terminating condition of the SPAIN algorithm.
In each step, a shortest path is computed and
the cost of the path edges is increased by a constant large number 
\cite{Mudigonda2010},
in order to prevent their use in the following iterations,
until $k$ different paths have been found or 
the new path has already been computed in a previous iteration.
By using a deterministic shortest path algorithm 
(such as Dijkstra's algorithm),
a path can be computed twice
before all distinct shortest paths have been found,
in which case the SPAIN algorithm stops prematurely
and returns less than $k$ paths. 
The authors also recognize this limitation 
(in \cite{Mudigonda2009}, section 8.3).\footnote{To see an example,
let $x$ be the leftmost leaf and $y$ be the rightmost leaf 
of the two level hierarchical network 
(depicted in Figure~\ref{fig:fattree}).
The first four paths found have the form:
$x\, v_1\, a\, w_1\, y$,
$x\, v_2\, a\, w_2\, y$,
$x\, v_1\, b\, w_1\, y$ and
$x\, v_2\, b\, w_2\, y$,
where
$v_1$ and $v_2$ are the nodes adjacent to $x$, 
$w_1$ and $w_2$ are the nodes adjacent to $y$, and
$a$ and $b$ are the upper level nodes.
If the edge cost increment is 24 
(because there are 24 edges of cost 1 in the original network),
at this moment of the execution, 
the edges incident upon $x$ or $y$ have weight 49 ($1+24+24$) and 
those incident upon $a$ or $b$ have weight 25. 
So, the 8 original shortest paths have the same cost again.
Consequently, the fifth computed path is equal to the first one
and the algorithm stops.}

Computing the best paths for the three level hierarchical network, 
the most long computation with both algorithms, 
took 4722 milliseconds with Algorithm~1  
and approximately 181 milliseconds with the SPAIN algorithm.

\subsubsection*{Backbone networks.}

In these networks all POP nodes are considered network edge nodes.
For each pair of edge nodes, 
Algorithm~1 chooses paths among those obeying the user provided constraints
(set by $h$ and $f$),
which form a set called the \textit{search path set} or \textit{search set}.
Depending on the value of $k$ and the size of the search set,
the computation of a best set may have a non negligible cost.
Our experience shows that,
although the search set can be huge,
in general, 
very large search sets do not produce better results than smaller ones. 
Therefore,
we developed a new version of the algorithm (Algorithm~1'),
which allows for an extra parameter: 
an upper \textit{threshold} for the size of the search set.

At runtime, for each pair, 
Algorithm~1' starts by computing the search set
with the given values for $h$ and $f$.
When its size is over the user provided threshold, 
the values of $h$ and $f$ are automatically adjusted to shrink the search set.
Conversely, 
when the size of the search set is below $k$
and there are more paths, 
$h$ and $f$ are enlarged.
Moreover, 
after selecting a best set (possibly with $k$ paths), 
if its disjointness degree is one,
an extra fully disjoint path is added 
unless it does not exist.
To compute this extra path the values of $h$ and $f$ are ignored. 
All those situations are flagged and logged.

Apart from selecting the set of paths, 
Algorithm~1' also computes several characteristics of
the selected sets (see below). 
Therefore, it is possible to run the algorithm using 
an increasing sequence of thresholds 
(\eg 100, 150, 200, 250, \dots).
As soon as there are no differences 
in the number of selected paths and
in the disjointness degrees, and 
the differences in the average path length and cost are below 
a certain bound (\eg 1\%), 
there is no need to augment the threshold.
The highest used thresholds are presented in Table~\ref{tab:select4}
and varied from 100, for the most simple network (B4), 
and 350, for the most complex ones (ATT and Sprint).

Table~\ref{tab:select1} presents an overview of the computed results 
for each backbone with $k=4$, $h=3$, $f=3$, and the mentioned thresholds.
The value $k=4$ is often referred to as adequate for a backbone, 
since greater values do not allow very significant improvements
in traffic distribution optimality 
(see for example \cite{heckmann2006,b4-sdn-wan}).
The first four columns of the table characterize 
the input network. 
The next three columns present 
the total number of paths computed and 
the averages of hop and latency stretches of these paths 
(an average of averages). 
Latency stretch is expressed in milliseconds 
in the same referential used to characterize link costs. 
The following three columns specify the percentage of pairs 
for which the computed set of paths
has disjointness degree 1, 2, or at least 3.
Finally,
the last column contains the number of pairs 
for which Algorithm~1' computed less than $k$ paths
although they existed in the network.

The need to resize a search set was quite common and 
the adjustments were automatically performed as explained above.
Whenever the algorithm found less than $k$ paths,
there were not $k$ paths between the two nodes in the network. 
This happened with 3 pairs in the Géant backbone and 
with 14 in the ATT backbone. 
In all other cases the algorithm found $k$ paths. 
Reflecting this, the last column of the table only has 0's.

\begin{table}
\caption{Results of backbone path computations with Algorithm~1'
for $k=4$, $h=3$, $f=3$ and the thresholds in Table~\ref{tab:select4}}
\label{tab:select1}
\begin{center}
\begin{tabular}{ |c|c|c|c|c|c|c|c|c|c|c| }
		\hline
		\multirow{3}{8ex}{Network} 
        & \#            
        & \#             
        & \#           
        & \# total  
        & Average      
        & Average 
        & \multicolumn{3}{c|}{\% pairs whose sets have}   
        & \# pairs                                         \\
	    & nodes         
	    & edges     
	    & pairs   
        & selected  
        & hop          
        & latency       
        & \multicolumn{3}{c|}{disjointness degree} 
        & w/ $<k$                                           \\ 
	    & $| V |$      
	    & $| E |$        
	    &                
        & paths   
        & stretch      
        & stretch    
        & \ \ 1 \  \ 
        & \ 2  \ 
        & 3 or 4  
        & paths                                             \\ 
		\hline
		\ Abovenet \ & 15 		& 30 		& 105 	& 420  	&   1.09	&  4.92  &    0     &    57.1  & 42.9  & 0 	\\
		ATT 			& 35 		& 68 		& 595 	& 2366  	&   1.21  &   5.00   & \  \ 16.1 \ \ &    \ \ 61.7 \ \  &   22.2 & 0  \\
		B4 			& 12 		& 19 		& 66	 	& 264  	&   1.28  &   12.9   &  0 &    87.9  &   12.1 & 0  \\
		Géant 		& 32 		& 49 		& 496 	& 1991  	&   1.75 	&  8.49  & 0  &    89.5   & 10.5  & 0 \\
		NTT 			& 27 		& 63 		& 351 	& 1404  	&  1.04	&  11.15   &   0     &    67.2  &  32.8  & 0   \\
		Sprint 		& 32 		& 64 		& 496 	& 1984  	&  1.25	&  6.49   &     0    &     66.1 &  33.9  & 0   \\
		Tiscali 		& 30 		& 76 		& 435 	& 1740  	&   0.81	&  1.69   &    0    &     53.1    &  46.9 &  0  \\
		\hline
\end{tabular}
\end{center}
\end{table}

Node pairs for which no two fully disjoint paths were computed are rare.
In the ATT backbone, 
after removing all degree 1 POPs, 
there are 3 POPs in Florida connected to the rest of the backbone 
by only one link, from Orlando to Atlanta. 
So, the 96 sets of paths that need to cross that link 
have disjointness degree 1. 
This characteristic of the network cannot be corrected
since we are using the topologies made available by the Rocketfuel project, 
which may not represent the physical network configuration.
Therefore, 
we only consider as worth noting situations 
where a 5th path was added to the 4 initially selected
to ensure disjointness. 
This occurred for 13 pairs in the Géant backbone and for 10 in the ATT.

Algorithm~1' is aimed at selecting $k$ paths (between a pair of nodes) 
that respect certain constraints relevant to their quality
(set by $h$ and $f$). 
Among the eligible paths, 
it selects a subset that maximizes edge disjointness. 
It is therefore interesting to compare its results with 
those of a greedy algorithm that only maximizes one of these facets.
For that purpose we chose again the SPAIN algorithm, 
whose results for the same backbones are presented 
in Table~\ref{tab:select2}.
Recall that the SPAIN algorithm mostly prioritizes edge disjointness and 
sometimes it is unable to find $k$ paths, even when they exist. 
The last column of Table~\ref{tab:select2} records those cases,
which were quite frequent.

\begin{table}
\caption{Results of backbone path computations with the SPAIN algorithm
for $k=4$}
\label{tab:select2}
\begin{center}
\begin{tabular}{ |c|c|c|c|c|c|c|c|c|c|c| }
		\hline
		\multirow{3}{8ex}{Network} 
        & \#            
        & \#             
        & \#           
        & \# total  
        & Average      
        & Average 
        & \multicolumn{3}{c|}{\% pairs whose sets have}   
        & \# pairs                                         \\
	    & nodes         
	    & edges     
	    & pairs   
        & selected  
        & hop          
        & latency       
        & \multicolumn{3}{c|}{disjointness degree} 
        & w/ $<k$                                           \\ 
	    & $| V |$      
	    & $| E |$        
	    &                
        & paths   
        & stretch      
        & stretch    
        & 1         
        & 2  
        & 3 or 4  
        & paths                                             \\ 
		\hline
		\ Abovenet \	& 15 		& 30 		& 105 	& 385  	&   0.98	&  4.87 & 	0     &    57.1  & 42.9  &  	22 	\\

		ATT 		& 35 		& 68 		& 595 	& 2231  	&   1.27  &   5.39   &	\ \ 16.1 \ \  &  61.0  &   22.9 & 	91 	 \\
		B4 		& 12 		& 19 		& 66 		& 227  	&   1.13  &   13.86   &	0	  &  78.8  &   21.2 & 	21 	 \\		
		Géant 	& 32 		& 49 		& 496 	& 1796  	&   2.30 	&  8.64  &	0     &    \ \ 84.3 \ \   &  \  15.7 \  & 	107	 \\
		NTT 		& 27 		& 63 		& 351 	& 1321  	&  1.11	&  9.85   & 	0     &    65.8  &  34.2  & 	43  	\\
		Sprint 	& 32 		& 64 		& 496 	& 1878  	&  1.49	&  8.42   &	0    &     64.7 &  35.3  & 		74   	\\
		Tiscali 	& 30 		& 76 		& 435 	& 1557  	&   0.83	&  1.68   &	0    &     51.8    &  48.2 &  	140 	 \\
		\hline
\end{tabular}
\end{center}
\end{table}

The results show that 
Algorithm~1' computed $k$ (or $k+1$) paths in all cases,
as long as they existed,
whereas the SPAIN algorithm did not achieve the main goal frequently,
for all backbones.
There is no remedy for this problem,
which becomes more serious with higher values of $k$.
In spite of the SPAIN algorithm privileging edge disjointness,
the results show that it only covers a higher number of faults 
in a relatively low percentage of pairs.

This trend is understandable and expected. 
We carried out several experiments with these networks, 
using the same values of $k$ and $h$
while changing the value of $f$. 
Increasing $f$, enlarges the search set, 
what often resulted, not only in bigger disjointness degrees,
but also in greater cost and hop stretches.
This also illustrates the flexibility Algorithm~1 provides.

Finally, Table~\ref{tab:select4} presents 
the thresholds used,
the total execution time required by Algorithm~1'
to compute all paths, as well as the average per pair of nodes.
As the time spent with each pair is highly variable, 
even with search sets of similar size, 
the table also contains the worst case, 
\ie the maximum execution time for a single pair.
Each pair computation being independent, 
the program can be easily speedup by parallelization. 
By contrast, the SPAIN algorithm required 617 milliseconds
to compute all sets of paths for the ATT backbone,
which turned out to be its more demanding network.

\begin{table}
\caption{Thresholds and execution times (in seconds) for Algorithm~1'}
\label{tab:select4}
\begin{center}
\begin{tabular}{ |c|c|c|c|c|c|c|c|c|c|c|}
		\hline
		\multirow{2}{8ex}{Network} 
		& \multirow{2}{8ex}{ \ \# pairs \ }
		& \multirow{2}{12ex}{ \ Threshold \ } 
		& \multicolumn{3}{c|}{Execution time} \\
		&  & 
		& \ all pairs \ 
		& \ average per pair \ 
		& \ worst pair \                      \\
		\hline
		\ Abovenet \ & 105 & 150 & 8.50	&  0.081  & 0.692 \\
		ATT         & 595  & 350 & 372 	&  0.625 & 32.495 \\
		B4          & 66  & 100 & 1.46	&  0.022 & 0.194 \\
		Géant      & 496 & 200 &  67.7 &  0.136 &  5.384\\
		NTT        & 351  & 250  & 24.3 	& 0.068 & 2.102 \\
		Sprint      & 496  & 350  & 1240 &  2.50 & 43.618 \\
		Tiscali      & 435 & 250 & 10.6 &  0.024 & 0.539 \\
		\hline
\end{tabular}
\end{center}
\end{table}

\subsection{Conclusions concerning the path selection algorithm}

Algorithm~1 is not polynomial and, consequently, 
its running times are greater than those of greedy algorithms.
For each pair of nodes, $(x,y)$, 
it searches a set of $k$ paths connecting $x$ to $y$ 
that maximizes disjointness, 
among all such $k$ paths sets that respect user given
hop and cost (or latency) stretch constraints.

Algorithm~1' allows a quite fine-tuning of 
the paths from which the best set is selected,
using parameters $h$, $f$ and the threshold on the search set size,
what is singular among algorithms with similar aims.  
Naturally, 
all this flexibility comes at the cost of an increase in running time. 
This increase has been shown as manageable and 
can be reduced using parallel execution.
Also, recall that this algorithm is only executed off-line
and not frequently.
Besides, it also computes some properties of the chosen paths and 
dynamically corrects the most striking anomalies, 
caused by the user given parameters
or the characteristics of the network. 
All in all, it allows 
a deeper knowledge of the network topology and 
a close control of the features of the computed set.

We now turn to the problem of efficiently using the computed paths 
to materialize concrete routing strategies in the network 
by aggregating them into trees.


\section{Path aggregation into trees}
\label{sec:trees}

As it is evident from the previous section, 
and clearly highlighted by the total number of paths computed 
(see Table~\ref{tab:select1}), 
multi-path routing for traffic engineering requires 
the usage of a very significant number of different paths, 
of order $k n^2$.
Moreover, 
different classes of traffic may still increase path numbers.

If one wants to avoid complex and dynamic routing algorithms 
in the core of the network, 
these paths must be mostly pre-configured,
and reducing the state required to setup them 
(\ie the size of the FIBs in routers) is an important goal. 
As it will be presented in section~\ref{sec:tree-routing}, 
using trees for this aggregation is one of the best solutions. 
Unfortunately,
determining the minimum number of trees needed to cover a set of paths is 
an NP-hard problem.

\subsection{The problem and known solutions}
\label{sec:trees-related}

Given a set $S$ of simple paths in 
an undirected, simple and connected graph $G$, 
the \textit{aggregation of paths into trees problem} consists in computing 
a set $T$ of trees,
with minimum cardinality,
such that each path in $S$ is a path in some tree of $T$.
Recall that a tree is 
an undirected, simple and connected graph 
that is acyclic.\footnote{A cycle 
is a path with at least two nodes, 
where the first and the last nodes are equal and 
whose edges are all different.}
Without loss of generality, 
we assume that every path $p \in S$ has at least two nodes.
There is a subproblem of ours,
named \textit{aggregation of paths with the same destination into trees}, 
where all paths in $S$ have the same destination.

Deciding if a set of paths with the same destination
can be aggregated into $m \geq 1$ trees is an NP-complete problem
\cite{Bhatnagar2002,Mudigonda2009}. 
Therefore, 
both optimization problems referred to above are NP-hard and
the known polynomial algorithms for solving them do not guarantee 
that the computed set of trees has minimum cardinality.
The size of the returned set is the main metric to evaluate their quality.

The subproblem of aggregating paths with the same destination
has been studied in the context of computing MPLS LSPs m-t-p.
In \cite{lsps} it is formulated as 
a 0--1 integer linear programming problem. 
A greedy algorithm is proposed in \cite{Bhatnagar2002}, 
which basically aggregates paths (and trees) in decreasing order
of the length of their longest ``common suffixes''.



Although the most general problem could be tackled with those algorithms, 
the final number of trees would be far from a reduced one. 
In general, 
$S$ has $k$ paths for each pair $(x,y)\in N^2$, with $x < y$.
Thus, 
$|S| \approx k\frac{n(n-1)}{2}$, where $n=|N|$.
Partitioning paths in $S$ by the destination node 
would give rise to $n-1$ aggregation problems with the same destination and, 
for each one, 
the minimum number of computed trees would be $k$ because 
each of the $k$ paths with the same origin (and destination) 
would have to be covered by a different tree. 
Therefore, 
the final total number of trees would be at least $(n-1)k$, 
repetitions being not expected. 
This strategy will be dubbed \textit{strategy LSPs m-t-p} 
in section~\ref{sec:trees-evaluation}.


To simplify the discussion, 
let us consider that a path
$p=v_1 v_2 \cdots v_{m-1} v_m$ (with $m\geq 2$)
\textit{induces} the undirected graph $G_p = (V_p, E_p)$,
where $V_p=\{v_1, v_2, \ldots, v_{m-1}, v_m\}$ is the set of nodes and
$E_p=\{(v_1, v_2), \ldots, (v_{m-1}, v_m)\}$ is the set of edges,
\ie $G_p$ has the nodes and the edges in $p$.
Let also $G'=(V',E')$ be an acyclic subgraph of $G$ and $p \in S$.
$G'$ is said to \textit{contain} or \textit{cover} $p$ 
if $E_p \subseteq E'$, and
$p$ is \textit{aggregable} into $G'$ 
if $(V' \cup V_p, E' \cup E_p)$ is an acyclic graph. 
The \textit{insertion} or \textit{aggregation} of $p$ into $G'$
transforms $G'$ into the graph $(V' \cup V_p, E' \cup E_p)$.


In the context of system SPAIN \cite{Mudigonda2009,Mudigonda2010},
whose goals are similar to ours,
Mudigonda \etal developed two randomized algorithms
for aggregating paths into acyclic subgraphs (not necessarily connected).
Due to their randomized nature, 
both algorithms must be executed several times, 
being returned a computed set of subgraphs with minimum size.

The first algorithm is quite simple \cite{Mudigonda2009,Mudigonda2010}. 
Initially,
the set $R$ of subgraphs is empty. 
Set $S$ is traversed randomly and each of its paths, $p$, 
is treated sequentially,
by testing if it is covered by some subgraph in $R$.
If that is the case, $p$ is skipped; 
otherwise, $R$ is traversed again, in a random order,
up to find a subgraph $G'$ into which $p$ can be aggregated
and $p$ is inserted into $G'$. 
When no such subgraph is found,
graph $G_p$ becomes a new member of $R$.


The second algorithm \cite{Mudigonda2009} is much more complex
and its full description is outside the scope of this paper. 
Essentially, the algorithm has two phases. 
In the first one, 
$S$ is partitioned by the path destination node and
each subproblem is solved by a reduction to the vertex colouring problem.
Since the union of all computed sets 
(which is a solution of the original problem) 
can have a large number of subgraphs,
the second phase tries to merge subgraphs
in order to reduce the size of the returned set.
The motivation presented by the authors to design this algorithm 
is the parallel execution of the first phase
(due to the independence of the subproblems).
However, the paper does not contain any comparison of both algorithms
from the point of view of performance or the quality of the solutions. 
For this reason, 
our experiments include only the first algorithm 
whose implementation is simpler.


%

\subsection{Algorithm 2 -- the path aggregation algorithm}
\label{sec:trees-algorithm}


Algorithm~2 aggregates paths into trees in a deterministic way.
Starting from an empty set of trees,
paths are successively aggregated into the current trees, 
creating a new tree 
only when their insertion into any existing tree 
would result in a cyclic or disconnected graph. 
The prime difference towards the similar algorithms presented above
is that ``pairs of compatible paths'' are first inserted 
in a specific order and into specific trees. 

With some abuse of notation, 
a set $\{p,q\}$ with two distinct paths is called a \textit{pair}, $(p,q)$,
in spite of the order irrelevance. 
Recall that: 
the input set, $S$, is a set of paths in a graph $G$; 
$G_p = (V_p,E_p)$ denotes the graph induced by a path $p \in S$; and
a tree $t$ of $G$ is a connected and acyclic subgraph of $G$ 
(defined by $(V_t,E_t)$).

The key notion of ``compatibility'' is defined 
over two paths, 
over a path and a tree, and 
over a pair of paths and a tree. 
Compatibility of two paths will be used to specify the order 
in which paths pairs are processed.
The two other compatibility types will be used to identify
the tree where a path or a pair of paths is inserted,
when there are several alternatives.

By definition, 
the \textit{compatibility degree} 
of a pair of paths $(p,q)$ is $-1$,
if graph $(V_p \cup V_q, E_p \cup E_q)$ is cyclic; 
otherwise, it is $|V_p \cap V_q|$, the number of common nodes.
The \textit{compatibility degree} 
of a path $p$ and a tree $t$ 
is defined in an identical way: 
it is $-1$ if graph $(V_t \cup V_p, E_t \cup E_p)$ is cyclic; 
and $|V_t \cap V_p|$, otherwise.
The \textit{compatibility degree} 
of a pair of paths $(p,q)$ and a tree $t$ is:
$-1$ if graph $(V_t \cup V_p \cup V_q, E_t \cup E_p \cup E_q)$ is cyclic;
and $|V_t \cap V_p| + |V_t \cap V_q|$, otherwise. 

Two paths (respect., a path and a tree, or a pair of paths and a tree)
are said to be \textit{compatible} if their compatibility degree
--- denoted by $\compat(\cdot,\cdot)$ --- is positive.
Note that two paths (respect., a path and a tree) without common nodes 
are not compatible,
because the union of the corresponding graphs is disconnected.
However, a pair of paths can be compatible with a tree 
even if one of the paths does not share any node with the tree
(providing the other does).

The following properties,
which justify the operations on compatible entities,
are easy to verify:

\begin{itemize}
\item If $(p,q)$ is a pair of compatible paths, 
      the graph $(V_p \cup V_q, E_p \cup E_q)$ 
      \textit{created} with $(p,q)$ is a tree.
\item If a path $p$ and a tree $t$ are compatible,
      the \textit{insertion} of $p$ into $t$, 
      which transforms $t$ into 
      the graph $(V_t \cup V_p, E_t \cup E_p)$,
      yields a tree.
\item If $(p,q)$ is a pair of compatible paths,
         $t$ is a tree, and 
         $(p,q)$ and $t$ are compatible,
      the \textit{insertion} of $(p,q)$ into $t$, 
      which transforms $t$ into the graph
      $(V_t \cup V_p \cup V_q, E_t \cup E_p \cup E_q)$,
      yields a tree.
\end{itemize}

The algorithm comprises four main phases:
(i)   generation of all pairs of compatible paths;
(ii)  aggregation of pairs of compatible paths;
(iii) generation of all \textit{single} paths,
      which are the paths in $S$ that were not treated in the previous step 
      (because, for instance, they are incompatible with any other path);
(iv)  aggregation of single paths. 

In the first phase, 
the compatibility degrees of all pairs of paths in $S$ are computed.
Compatible pairs will be processed in the second phase, 
in an order that follows three criteria: 
first, in decreasing compatibility degree of the pair;
second, in decreasing ``aggregation potential'' of the pair in $S$; and,
in case of equal compatibility degrees and aggregation potentials, 
in decreasing length of the pair 
(which is the sum of the lengths of both paths).

The first criterion aims at privileging the pairs 
whose aggregation is more ``natural'', 
\ie those whose aggregation result differs less from each of the paths. 
The ``aggregation potential'' gives priority to pairs of paths 
that share many common nodes with their compatible paths.
Formally, 
the \textit{aggregation potential} of a path $p$ in set $S$ is 
the sum of the compatibility degrees of all compatible pairs of $S$ encompassing $p$:
$$\potencial(p,S) = 
  \sum_{\{ q \in S \; \mid \; 
          q \neq p \; \wedge \; \mbox{\scriptsize compat}(p,q) > 0\}} 
  \compat(p,q).$$
The  \textit{aggregation potential} of a pair of paths $(p,q)$ in set $S$ is
the sum of the \textit{aggregation potentials} of $p$ and $q$ in $S$:
$$\potencial((p,q),S) = \potencial(p,S) + \potencial(q,S).$$
Finally, the goal of the third criterion is 
to treat the longest paths as soon as possible, 
when there are more alternatives, 
deferring those that, in principle, are easier to aggregate.

In the aggregation phase of the compatible pairs of paths, 
each pair is processed in the order introduced above, 
until there are no more pairs left. 
The algorithm starts by verifying, for each path of the pair, 
whether it is covered by some of the existing trees. 
Three cases can arise. 
In the first one, 
both paths are contained in trees (possibly different) and
the processing of the pair ends. 
In the second case, 
none of the paths is contained in a tree. 
If there is some tree compatible with the pair, 
the pair is inserted into an existing tree (specifyed below);
otherwise, a new tree is created with the pair.
In the third case, 
one of the paths is contained in some tree $t$ and
the other (which will be designated by $p$) is not covered by any tree. 
If $p$ is compatible with $t$, 
it is aggregated into $t$; 
if $p$ is not compatible with $t$ but 
there is some tree compatible with $p$, 
$p$ is inserted into an existing tree (see below); 
otherwise, no tree is compatible with $p$, 
so $p$ is ignored and 
its aggregation is postponed until the fourth phase.

In the third phase,
paths that have not been treated yet are sorted 
in decreasing order of length,
to be processed in the last step.
The processing of a singular path $p$ also starts
by searching a tree that covers it.
If there is any, nothing more is done.
Otherwise, if $p$ is compatible with some tree,
$p$ is inserted into an existing tree;
when $p$ is incompatible with all available trees, 
tree $G_p$ is created and added into the result set. 

The search for a compatible tree with a path or a pair of paths $\alpha$
returns, in case of success, 
the tree $t$ where $\alpha$ will be inserted. 
That tree is, among all those compatible with $\alpha$,
one that maximizes the compatibility degree. 
Thus, if $T$ is the set of all existent trees, 
the tree $t \in T$ where $\alpha$ is aggregated verifies:
$\compat(\alpha,t) \geq 1$ and
$(\forall t'\in T)\; \compat(\alpha,t) \geq \compat(\alpha,t')$.

The running time of the algorithm is $O(|S|^3 \times |V|^2)$,
where $V$ denotes the set of nodes of network $G$, 
due to the following.
The compatibility degree of two entities can be computed
in $O(|V|^2)$ steps,
because all nodes belong to $V$ and
the length of any path cannot exceed $|V|-1$.
Therefore, the first phase is $O(|S|^2 \times (|V|^2 + \log |S|))$,
since all pairs of paths are analysed and compatible pairs are sorted.
The third phase is $O(|S| \times \log |S|)$, due to sorting.
In the two aggregation phases, 
all pairs of compatible paths and all single paths are processed. 
As the processing of each pair or single path requires 
$O(|T| \times |V|^2)$ time, 
where $T$ represents the current set of trees,
the total cost of these phases is $O(|S|^2 \times |T| \times |V|^2)$.
The conclusion stems from the fact that $|T| \leq |S|$.
 
In the following section we present the evaluation of the algorithm 
while aggregating the paths computed by the path selection algorithm.

\subsection{Algorithm~2 evaluation}
\label{sec:trees-evaluation}

In this evaluation 
we used the sets of paths computed with Algorithms~1 and 1',
as presented in section~\ref{sec:selection-evaluation}. 
Once again, 
the tests performed with the synthetic regular networks 
(full mesh, ring, hierarchical and folded clos networks) 
were significant to assess the absolute quality of the results, 
since the minimum number of trees needed to cover all paths is known
(recall Table~\ref{tab:networks} in section~\ref{sec:regular-networks}).
Therefore, we can compare the results obtained with the optimal ones,
in contrast to those achieved with the backbone networks,
for which the optimal solutions are unknown.

Table~\ref{tab:tree-results} presents the most important data and results. 
The first columns identify the network, 
the number of edge nodes (n\ =\ $|N|$), 
the desired number of paths between each pair of nodes ($k$), 
and the total number of paths to be aggregated ($|S|$). 
For the regular networks, the minimum number of trees
required to cover the paths is then shown.
The next two columns contain
the number of trees built by Algorithm~2 and
the number of graphs computed by the first algorithm proposed 
for the system SPAIN 
(see \cite{Mudigonda2009,Mudigonda2010} and the discussion in
 section~\ref{sec:trees-related}).
The last column presents the number of trees that would be obtained
by the strategy LSPs m-t-p (see again section~\ref{sec:trees-related}),
which corresponds to $(n-1)k$.

\begin{table}
	\caption{Results of the path aggregation algorithms. 
	         The numbers of executions of the SPAIN algorithm are 
	         presented in Table~\ref{tab:tree-results2} and
	         its computing times correspond approximately to 100 times 
	         the time taken by Algorithm~2.}
	\label{tab:tree-results}
\begin{center}
\begin{tabular}{|c|c|c|c|c|c|c|c|}   \hline
		\multirow{2}{8ex}{Network} 
 		& \multirow{2}{9ex}{ \ $n = |N|$ \ } 
 		& \multirow{2}{10ex}{ \ \ \ \ \ \ \ $k$} 
 		& \multirow{2}{6ex}{ \ \ $| S |$} 
		& min
		& \multicolumn{2}{c|}{\# acyclic subgraphs} 
		& LSPs                                        \\
		& & &
		& \ \# trees \ 
		& \ Algorithm~2 \  
		& \ SPAIN \  
		& \ m-t-p \                                   \\ \hline
		full mesh      & 12 & 11 & 726 & 12 & 12   &  56   &   121 \\
		\hline
		ring           & 12 &  2 & 132 & 12 & 12 &  12   &    22   \\
		\hline
		hierarchical 2 &  8 &  2 or 8 & 152 & 8 & 8 & 10 & 56 \\
		\hline
		hierarchical 3 & 16 & \ 2, 8 or 32 \ & \ 2352 \ & 32 &   40  &  71  & 496 \\
		\hline
		folded clos 6  &  6 &  6 & 90 & 6 & 6 & 13 & 30 \\
		\hline
		\ folded clos 12 \ & 12 & 12 & 792 & 12 &  12  &   58  &132  \\
		\hline
		Abovenet  & 15 & 4 & 420 & --- & 24 & 26 & 56 \\
		\hline
		ATT & 35 & 4 & 2366 & --- & 54 & 65 & 136 \\
		\hline
		B4  & 12 & 4 & 264 & --- & 20 & 20 & 44 \\
		\hline
		Géant       & 32 & 4 & 1991 & --- & 57 & 69 & 120 \\
		\hline
		NTT         & 27 & 4 & 1404 & --- & 42 & 56 & 104 \\
		\hline
		Sprint  & 32 & 4 & 1984 & --- & 52 & 68 & 124 \\
		\hline
		Tiscali  & 30 & 4 & 1740 & --- & 27 & 36 & 116 \\
		\hline

\end{tabular}
\end{center}
\end{table}

Being a randomized algorithm, 
the SPAIN algorithm was executed several times with the same set of paths. 
But, instead of defining the exact number of executions 
(also called iterations) per network, 
we limited its total execution time to 100 times 
the time required by Algorithm~2 for the same network.
Table~\ref{tab:tree-results} has the minimum number of subgraphs 
computed in all iterations. 
Table~\ref{tab:tree-results2} presents the execution times of Algorithm~2,
the thresholds for the total execution time of the SPAIN algorithm and 
the number of iterations it actually performed during that time.
It is worth noting that this algorithm has been implemented
as described in section~\ref{sec:trees-related} and 
by the authors of \cite{Mudigonda2009,Mudigonda2010}.
Consequently, 
the output is a set of acyclic graphs, not necessarily connected 
(what justifies the common title of the two 
Table~\ref{tab:tree-results} columns that present the results).

\begin{table}
	\caption{Execution times (in seconds) of Algorithm~2,
	         thresholds (in seconds) for the total execution time 
	         of the SPAIN algorithm
	         and corresponding numbers of iterations performed}
	\label{tab:tree-results2}
\begin{center}
\begin{tabular}{ |c|c|c|c|c|c|c|}
\hline
		\multirow{2}{8ex}{Network} 
        & Algorithm~2 
        & \multicolumn{2}{c|}{SPAIN}                \\
        & \ execution time \     
        & \ threshold \    
        & \ \# iterations \                          \\
		\hline
	full mesh   &     0.197   &      19.7   &   4384  \\
		\hline
	ring        &     0.002   &       0.2   &   2140  \\
		\hline
	hierarchical 2 &  0.034   &       3.4   &  10417  \\
		\hline
	hierarchical 3  & 16.565  &    1656.5   &  71642  \\
		\hline
	folded clos 6   &  0.012  &       1.2   &   3270  \\
		\hline
	\ folded clos 12 \ & 0.302&      30.2   &   2793  \\
		\hline
	Abovenet        & 0.067   &       6.7   &   6169  \\
		\hline
	ATT    &   2.276   &  227.6  &   14125  \\
	    \hline
	B4     &  0.020  &   2   &  3247  \\
		\hline
	Géant     &   1.091  &   109.1  &   7533  \\
		\hline
	NTT     &   0.670 &       67   &    7623  \\
		\hline
	Sprint      &  1.159  &       115.9   &   8671  \\
		\hline
	Tiscali   &   1.885   &   188.5 &   33978  \\
		\hline

\end{tabular}
\end{center}
\end{table}

Results show that Algorithm~2 computes the optimal solution 
for almost all regular networks but the hierarchical 3 one,
where the result is approximately 25\% far from the minimum. 
In spite of the allowed running times,
the SPAIN algorithm only discovered the optimal trees
with the ring network.
In all other cases it returned a number of graphs 
between 37.5\% (hierarchical 2) and 383\% (folded clos 12)
far from the optimal.
In what concerns the backbones, 
where optimal solutions are not known,
Algorithm~2 always outperformed the SPAIN algorithm,
except in the B4 case for which both returned equivalent sets.
Values in the LSPs m-t-p column are always substantially higher than 
those obtained with the two algorithms,
which indicates a lack of effectivity in that strategy 
when the goal is to significantly reduce the size of the FIBs 
in the core of the network.
We will discuss in the next section how these results can be used to 
drive traffic routing in the network.

\section{Implementing multi-path routing 
using trees and off-the-shelf network equipment}
\label{sec:tree-routing}


In the previous two sections we have presented and evaluated two
algorithms intended to support
multi-path routing in a backbone network.

The first algorithm allows network managers to compute a
set of paths that can be used to route ingress flows to their
egress nodes, using load distribution among these paths. This paves the
way for offline or online optimization. We envision online
optimization using a centralized
controller that periodically adjusts traffic distributions.
We leave for future work the study of the relation between traffic 
distribution optimality and the estimated accuracy of the traffic matrix and 
of network path availability. We anticipate that it is possible to
find a reasonable solution even if the central controller only
updates its vision of network and the  traffic matrix from
time to time (with a period of several minutes for example).

The second algorithm supports a scalable way of path deployment in the core of 
the network by the way of a reduced number of trees. In the next paragraphs we will 
present several alternatives for this deployment using different
types of off-the-shelf equipment and simple protocols. 
By hypothesis, mechanisms required to implement traffic engineering
must be available in each ingress node $x$, allowing it to:
H1) given the destination address of a packet flow, to know $y$,
the flow egress node; H2) given an egress node $y$, to
know $k$ paths to get there; and H3) given the $k$ paths
that allow $x$ to reach $y$, which one should be chosen for
that flow (in accordance to some load distribution criteria).
The availability of these mechanisms is discussed below.
The way edge routers, servers or a central controller use them
to establish traffic engineering policies is left for future work.

\subsection{Routing with trees using MPLS and multipoint-to-point LSPs}

As already mentioned in the introduction,
MPLS is the most popular way to implement traffic engineering. 
Since the total number of different required LSPs may be very
important,  
\emph{multipoint-to-point} LSPs can be used
to reduce the state in the network
(see section~\ref{sec:trees}). In that section we have shown that it is possible
to have less state in the network by the way of leaving the
restriction that each such directed acyclic graph should be rooted in the egress node and
only supports traffic in the direction of the egress node.
Anyway, using MPLS for traffic engineering is a known
technique, and mechanisms H1),
H2) and H3) are part of the available implementations.

\subsection{Routing with trees using VLANs}

In a network of Ethernet switches it is possible to 
statically parametrize in each switch a mapping from
ports to VLANs \cite{Mudigonda2010}. Each tree, encompassing
different paths, is then supported by a different VLAN.
With this solution, the selection of a path to a destination switch (egress node) is
implemented by choosing the corresponding VLAN tag.
Routing is then performed using VLAN restricted
flooding and filtering on the basis of the destination address.

This solution has been proposed and tested in a data center network
scenario \cite{Mudigonda2010} and the required flow distribution
mechanisms (H1, H2 and H3) were implemented by the
servers Ethernet drivers. These drivers implemented a sort of static mapping from
server MAC addresses to switch addresses and VLAN tags, as well as
protocols to detect path availability and the speedup of
filtering mechanisms in the network switches. The authors of the
proposal used a uniform random distribution of the flows among the different 
available paths (or trees) and left for future work the study of the
impact of the usage of non uniform distributions.

\subsection{Routing with trees using longest-prefix matching routing}
\label{sec:lpm}

It is also possible to implement multi-path routing by the way
of  several rooted trees using hierarchical IP addresses and
\emph{Longest-Prefix Matching}. For this purpose, it is possible to parametrize routers,
using remotely setup static routing, in the following way. Each tree $t\in T$ is
associated with an IP prefix $I$ (an interval of addresses
which size is a power of 2). The intersection of all the
intervals must be empty. Then, each $I$ is partitioned in
as many sub prefixes as the root has descendants in this tree, and
each sub prefix of $I$ is associated with a different descendant node.
This process continues recursively up to the leaf nodes
of the tree. In each node, but the root, the parent
prefix is associated with all the interfaces leading to the
parent node, and each sub prefix is associated with all
the interfaces leading to the descendant associated with each sub prefix.
Finally, each node receives an address in its (sub)prefix
that it keeps for itself. Any node will have as many addresses
as trees it belongs to \cite{many-addresses}. The same applies for the number of
static routes associated with its interfaces. Each router
can also have an IP address, in a different prefix,
routed by a shortest-path protocol
to guarantee a direct control channel and a fallback path.

The choice of the path a packet will follow is performed
by choosing the address of the destination router in the tree
encompassing the chosen path. This packet address 
destination transformation can be performed by the way of
tunnels to preserve the original destination address.
Using LISP \cite{rfc6830} to support routing in this scenario
is a standardized way of implementing the tunnels, directly
providing interfaces to  mechanisms H1, H2 and H3.

\subsection{Routing with trees using OpenFlow}

According to several authors
(\eg \cite{sdn-internet}) and as already mentioned in the introduction, 
a vision where, for each different micro flow, state is installed in
the network is not scalable for intra-AS networking or for
intra data center routing. Also, recent proposals related to the usage 
of the SDN approach in the wide area intra-AS routing 
(\eg \cite{ms-sdn-wan,b4-sdn-wan}) rely on the usage of tunnels and
logically centralized controllers that perform path selection
decisions based on a global view of the network configuration
and status. 

Routing with trees, and their implementations based on VLANs or
on IP-based longest-prefix matching routing, can also be
implemented in the core of the network using OpenFlow.
Their usage requires the evaluation of a tradeoff between routing space and
forwarding complexity.

%
%

\section{Discussion and conclusions}
\label{sec:conclusions}

Provider backbone networks need flexible and adaptable
mechanisms to support traffic engineering with varying
demands and promptly adaptation to network faults. 
Currently, these goals are fulfilled using expensive and sophisticated
equipments that run distributed algorithms and implement dynamic routing in the
core of the network (\eg MPLS tunnels and dynamic trunk rerouting) as well as
a certain degree of over provisioning the network.

One of the most long standing principles of complex systems design,
the end-to-end argument, argues for a core as simple as possible and
relegates to the edge most of the complexity and adaptation to varying
 requirements.
Applied to routing, this design rule would avoid as much as possible
complexity in the core and, in particular, avoid the implementation of the support of
diversity and adaptation to varying requirements in the core, 
as currently is partially the rule. 
This line of thinking is aligned  with proposals
like those presented in \cite{drrch,Mudigonda2010,Suchara2011}
and is also partially aligned with the SDN
proposal \cite{OpenFlow,road-to-sdn,sdn-internet}.

Our quest for a simpler support for traffic engineering without
sacrificing optimality and adaptability to demand variations and
network faults lead us to envision a simple core, able to a priori provide
as many paths as required by the edge to adapt traffic to variable
requirements, as proposed in \cite{Mudigonda2010} for example,
or to network faults as proposed in \cite{Suchara2011} for example.

To pre-compute these path sets we developed Algorithm~1 for
path selection in a fault and demand agnostic way.
Its evaluation, in the context of several provider backbones,
showed that it is feasible to pre-compute a set of paths able to
support continuous connectivity and providing sufficient diversity
 to support traffic rerouting in face of a set faults. We insure 
 that these sets support at least one faulty link, what seems compatible
 with current practice if there is no lower level implied correlation
 among faults, like several links sharing the same fiber or the
 same conduct.  Altough the full
evaluation of the ability of the computed path sets to support flexible adaptation to
traffic variability is left for future work, many authors have 
referred that using 3 to 5 different paths between each pair of
edge nodes is sufficient to support almost optimal traffic load balancing
when path link disjunction is maximized as we also do.
Algorithm~1 is more complex than other alternatives but,
as we have shown, it always finds the demanded number of paths
if they are available in the network, and the computed set is also able to 
support at least one link fault if the network also is.

The total number of computed paths is naturally very important and
their continuous availability  requires many FIB entries in
core routers. To address this problem we have developed 
Algorithm~2, for aggregating paths into trees, with better results than
the previously proposed algorithms with the same goal.
Finally, we have shown that  off-the-shelf equipment supporting
simple protocols may be used to implement routing with a
reduced number of trees, what shows that simplicity can be
achieved by using only trivially available protocols and their most
common and unsophisticated implementations.

In networks where providers control not only the network but also customer
demands, like in private inter data center networks, 
an architecture is emerging \cite{b4-sdn-wan,ms-sdn-wan,multi-class}: 
a centralized controller (or a hierarchy of
several ones) receives information on traffic demands, schedules
them, and adapts routing decisions to optimize network
usage. The ultimate goal being to avoid as much as possible
network over provision.
This kind of control architecture can only be partially adopted in
a provider network since the provider has almost no control over
customer demands. However, we think that it is possible to adopt
a similar architecture. To make it real, there are several problems that must be
addressed: how can the central controller timely know current customer
demands and influence them? how can the central controller 
dynamically adapt routing to demands? what is the cost versus accuracy
needed to close the loop on customer demands awareness and
optimality of traffic engineering? and so on.

We intend to tackle these problems in our future work. The
algorithms introduced in this paper are central to the conception of that
architecture, since they permit a network design where in fact all
complexity is concentrated in edge routers and the controller,
while the core can be simple and based on off-the-shelf equipment
and off-the-shelf simple protocols.

\bibliographystyle{abbrv}

{\small

\bibliography{bibliography,rfc}
}

\end{document}